\documentclass[doublecol]{epl2} 
\usepackage{graphicx}
\newcommand{\dzt}{$d_{3z^2-r^2}$}
\newcommand{\dxtyt}{$d_{x^2-y^2}$} 
\title{Orbital control in strained ultra-thin LaNiO$_3$/LaAlO$_3$ superlattices}
\shorttitle{Orbital control in strained ultra-thin LaNiO$_3$/LaAlO$_3$ superlattices} 

\author{J. W. Freeland\inst{1}, Jian Liu\inst{2}, M. Kareev\inst{2}, B. Gray\inst{2}, J.W. Kim\inst{1}, P. Ryan\inst{1}, R. Pentcheva\inst{3}, and J. Chakhalian\inst{2}}

\shortauthor{F. Author \etal}

\institute{                    
  \inst{1} Advanced Photon Source, Argonne National Laboratory, Argonne, IL 60439\\
  \inst{2} Department of Physics, University of Arkansas, Fayetteville, AR 72701\\
  \inst{3} Department of Earth and Environmental Sciences and Center of Nanoscience (CENS), University of Munich, Theresienstr. 41, 80333 Munich, Germany
}
\pacs{nn.mm.xx}{Third pacs description}

\abstract{
In pursuit of rational control of orbital polarization, we present a combined experimental and theoretical study of  single unit cell superlattices of the correlated metal LaNiO$_3$ and the band insulator LaAlO$_3$. Polarized x-ray absorption spectra show a distinct asymmetry in the orbital response under strain. A splitting of orbital energies consistent with octahedral distortions is found for the case of compressive strain. In sharp contrast, for tensile strain, no splitting is found although a strong orbital polarization is present.  Density functional theory calculations including a Hubbard U term reveal that this asymmetry is a result of the interplay of strain and confinement induces octahedral rotations and distortions and altered covalency in the bonding across the interfacial Ni-O-Al apical oxygen, leading to a charge disporportionation at the Ni sites for tensile strain.}

\begin{document}
\maketitle

The interplay between structure and strong correlations in transition metal (TM) oxides gives rise to a breadth of intriguing phenomena ranging from high-temperature superconductivity to magnetism\cite{orbrev}. The ground state properties are fundamentally influenced by the orbital occupation. For the case of charge and spin degrees of freedom, electric and magnetic fields provide a direct means to control the energy of those elements of the ground state. For the case of orbital degrees of freedom the route to direct control is not well established. With the recent advances in oxide film synthesis, strain fields in perovskite heterostructures are by far the most common route used to attempt rational control of orbital energies\cite{orbrev}. However, recent experiments have demonstrated that altered chemical coordination at the interface can lead to a reconstruction of the orbital configuration into a unique correlated electron states not found in the bulk\cite{jcscience,laosto,smadici,jwfcrocmo,pyu}.

It was recently theoretically proposed that through a combination of strain and spatial confinement  one can possibly manipulate the orbital state in heterostructured LaNiO$_3$ (LNO). In bulk, LaNiO$_3$, the Ni$^{3+}$ (3d$^7$, $t^6_{2g}e^1_g, S = 1/2$) forms a paramagnetic metallic state\cite{goodenough2} with no evidence of orbital preference\cite{urs}, but it was theoretically suggested that strain and confinement could break this symmetry in  LaNiO$_3$/LaMO$_3$ (M=Al,Ga,...) superlattices and thereby allow the manipulation of the orbital occupancy\cite{lnolaoth1}. Calculations within the local density approximation in combination with dynamical mean theory  supported the formation of a cuprate-like Fermi surface in the LaNiO$_3$/LaMO$_3$ system\cite{lnolaoth2}. Additional theoretical work also demonstrated that the chemical confinement is more important than strain in controlling the orbital polarization\cite{millis}. On the experimental side, several recent studies have looked at several unit cell superlattices and shown a metal to insulator transition due to confinement effects\cite{sun, liuprb}, which was attributed to onset of a charge-ordered antiferromagnetic state at low temperature\cite{boris}. Polarization dependent resonant reflectivity was also utilized to demonstrate an enhanced orbital polarization at the interface with respect to the bulk of the LNO layer\cite{Benckiser}. However, these studies are all limited to tensile strain and a full understanding requires a microscopic picture in the ultrathin limit that can deconvolve effects of strain vs.\ confinement.

In this letter, we present a combined experimental and theoretical study on single-unit-cell LaNiO$_3$/LaAlO$_3$ superlattices to elucidate the possibility of cooperative chemical and strain control of orbitals. The results show an unexpected asymmetric orbital response to compressive vs. tensile strain. Compressive strain generates a small orbital splitting consistent with expectations of elastic theory while tensile strain shows no evidence of crystal field splitting yet displays a signifcant \dxtyt\ orbital polarization. Density functional theory (DFT) calculations with an on-site Coulomb repulsion term  confirm the distinct electronic behavior and attribute it to the concerted effects of broken symmetry at the interface, chemical confinement, and a strain induced lattice distortion.

Fully epitaxial [LaNiO$_3$(1 u.c.)/LaAlO$_3$ (1 u.c.)]$_{20}$ superlattices (SL) were grown on (001) TiO$_{2}$-terminated SrTiO$_3$ (STO) and LaAlO$_3$ (LAO) single crystal substrates by pulsed laser deposition with \textit{in situ} monitoring by RHEED using the interrupted growth method \cite{Blank, Kareev2, Kareev1}. The high-quality of the superlattice was confirmed by AFM and high-resolution x-ray diffraction at beamline 6-ID-B of the Advanced Photon Source (APS) at Argonne National Laboratory. To determine the electronic properties, the samples were studied with polarized x-ray absorption in the soft x-ray regime at beamline 4-ID-C of the APS. The spectra were measured in bulk sensitive fluorescence yield and aligned using a NiO (Ni$^{2+}$) standard measured simultaneously in the diagnostic section of the beamline.

\begin{figure}[h]
\begin{tabular}{r}
    \includegraphics[width=0.40\textwidth]{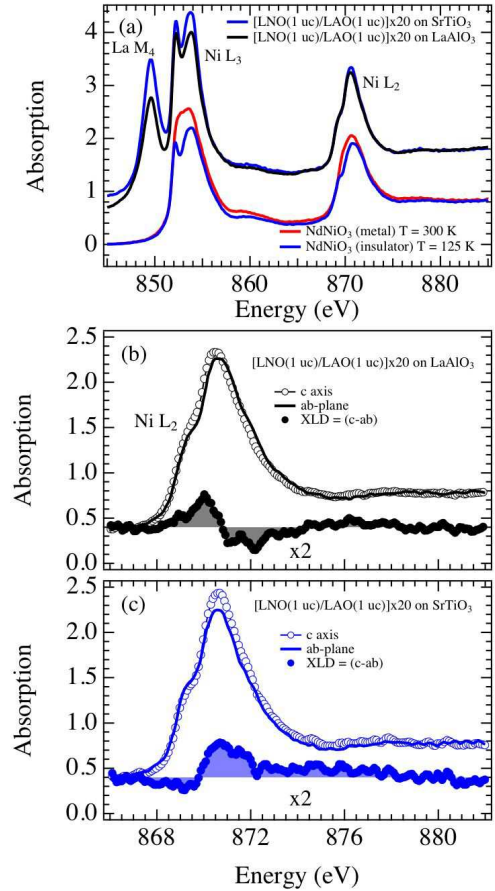} 
\end{tabular}
\caption{\label{xas}(a) Polarization averaged Ni L edge in comparison with NdNiO$_3$ above and below the metal to insulator transition. (b) Polarization dependent x-ray absorption at the Ni L$_2$ edge for both (b) compressive (LAO substrate) and (c) tensile (STO substrate) strain. The data show the expected response for compressive strain, but the the lack of energy splitting for tensile is quite unexpected. The differences in the orbital response are highlight in both cases by showing the x-ray linear dichroism (XLD). }
\end{figure}
X-ray diffraction was used to gain insight into the superlattice crystal structure. In a pseudo-cubic form, the  $R\bar{3}c$ bulk LNO has a lattice parameter of $\sim$3.83\ \AA. Off-specular mapping of the (222) reflection (not shown) determined that the films are fully lattice matched to the substrate for both cases. The average c-axis lattice parameter is 3.932\ \AA\ and 3.823\ \AA\ for samples on LAO and STO, respectively. The results for LAO are consistent with an out of plane expansion expected  for a volume conserving scenario. Surprisingly, despite the large tensile strain on the STO-substrate, the SL shows almost the bulk  LNO lattice parameter. This indicates that the superlattice structure is responding differently under tensile vs.\ compressive strain, which was seen in recent studies of LaNiO$_3$ films\cite{smay,jclno}. The superlattice peaks are observable with a half-order position, consistent with 2 u.c. periodicity of the superlattice\cite{Kareev1}.

\begin{figure}[h]\vspace{-0pt}
\begin{tabular}{c}
	\includegraphics[width=0.4\textwidth]{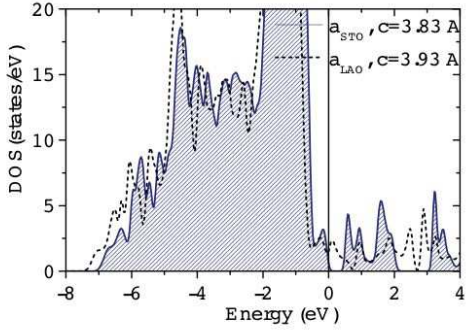}
\end{tabular}
\caption{\label{dost}Total density of states for (LAO)$_1$/(LNO)$_1$ for the case of tensile (STO) (solid black line) and compressive strain (LAO) (dashed blue line).}
\end{figure}
To  provide a more detailed insight into differences in the electronic properties as a function of strain we performed polarization dependent XAS at the Ni L edge . XAS probes the density of unoccupied states giving a picture of not only the chemical state, but also the orbital occupancy. An important point to note is that due to the strong core hole interaction, the Ni L-edge probes the local Ni environment. From the polarization averaged XAS, we determine that the valence of Ni is bulk-like 3+ testifying to  the absence of oxygen vacancies and lack of a significant amount of charge transfer at the interface (see Fig.\ \ref{xas}(a)). The data are shown in comparison to bulk NdNiO$_3$ above and below the metal-insulator transition. A strong multiplet feature develops on the leading edge of both the L$_3$ and L$_2$ edges, which is identified as a signature of localized carriers in insulating charge-ordered (CO) nickelates\cite{Piamonteze,jllno,jlnno}. The spectroscopy illustrates directly that, at a local level, the carriers are localized in agreement with the macroscopic scale results of the transport measurements (not shown). This finding of a charge ordered state is also consistent with recent optical studies of ultrathin superlattices\cite{boris}.

The effect of confinement and strain on orbital occupancies can be probed directly by light with tunable linear polarization. Due to the strong overlap between the La M$_4$ and Ni L$_3$ edges, however, for the polarization dependence we focus on the behavior of the L$_2$ edge only. First, consider the case of compressive strain (LAO) as shown in Fig.\ \ref{xas}(b). In-plane (ab) octahedral compression was shown with XRD to generate an out-of-plane (c) expansion of the NiO$_6$ ocathedra, which leads to a lowering of the c-axis orbital ($3z^2-r^2$) energy. As anticipated, this  effect is clearly observed in the polarized XAS, which shows a shift of the c-axis polarized spectra to lower energy as expected when lowering the energy of this state. This shift of $\sim$100 meV is related to the energy scale for the crystal field splitting due this type of structural distortion\cite{jclno}, but it is important to note that the crystal field splitting is small compared to the bulk LNO bandwidth of several eV\cite{goodenough2}. The polarization dependence is best seen in the difference of the ab- and c-axis spectra known as x-ray linear dichroism (XLD). Under compressive strain, the XLD shows a derivative like lineshape (see Fig.\ \ref{xas}(b)) consistent with the effect being largely an energy shift between the ab vs.\ c spectra. 

For the case of tensile strain, under the assumption of a symmetric strain response, one would expect the opposite effect, but the data in Fig.\ \ref{xas}(c) shows practically no energy splitting between the two orbitals. Instead, there is a significant difference in peak intensity with polarization. The lack of energy shift strongly implies there is no splitting of the $e_g$ band as expected for {\it undistorted} octahedra. On the other hand, even though there is no energy splitting of the orbitals under tensile strain, the difference in intensity signals a significant departure from the uniform occupation of \dzt\ vs \dxtyt\ orbitals in bulk LNO. In this case, the XLD shows almost no intensity in the region of the rising edge of the peak due to the absence of an energy shift, but has a large signal near the peak maximum due to the difference in $e_g$ orbital occupancies. 

\begin{figure}[h]\vspace{-0pt}
\includegraphics[width=0.48\textwidth]{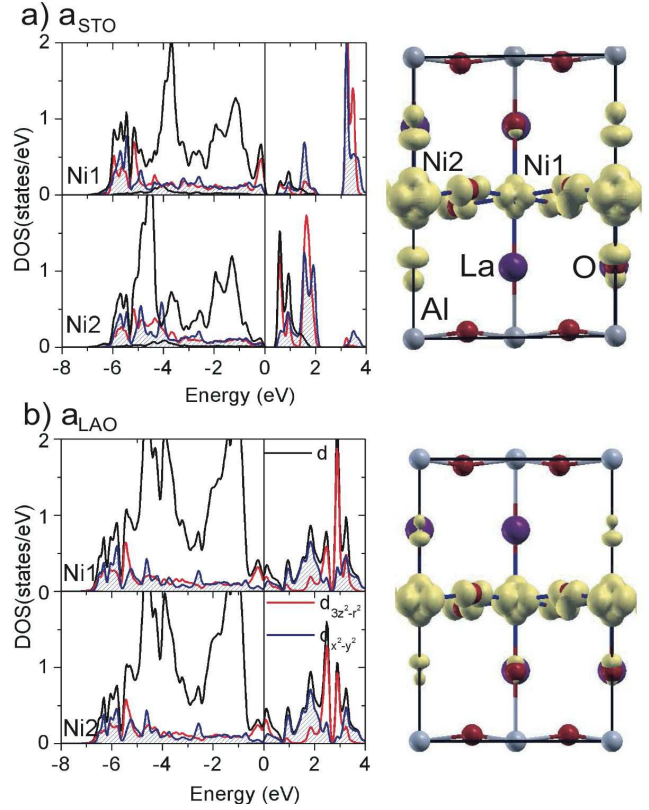}
\caption{\label{pdoscdn}  (left) Projected (spin-averaged) density of states (PDOS)   of Ni $3d$  and (right) electron density distribution of the unoccupied states ibetween E$_{\rm F}$ and 2.5 eV in (LAO)$_1$/(LNO)$_1$ for (a) tensile (STO)  and (b) compressive (LAO) strain. The right panels clearly show the  reduction in hole density on the apical oxygen between Ni and Al, as well as the charge disporportionation at the Ni sites for tensile strain.
}
\end{figure}
To quantify the differences in strain dependent orbital occupancy, we integrate the area under the peak to count the number of unoccupied states (holes) which scales with the orbital occupancies. Proper counting of holes requires measurement of the entire L edge, which is not possible due to the La M-edge intrusion into the L$_3$ edge. However, under the assumption that spin-orbit coupling in the $3d$ states is constant with strain, by integrating the L$_2$ edges we can extract the relative orbital occupancies\cite{spinorbitxas}. For both strain states the in-plane hole counts are very similar indicating that the \dxtyt\  orbitals have a  similar occupancy regardless of strain. For tensile strain though, there is a significant increase in the number of holes in the out-of-plance \dzt\ orbital states (decrease in electron occupancy of the \dzt\  band).  By defining orbital polarization through the relative areas, 
then there is  about 5\% reduction in \dzt\ occupancy for tensile strain, while for compressive strain the orbital polarization is close to zero. The polarization for tensile strain agrees well with  recent measurements using polarized x-ray reflectivity\cite{Benckiser}.

These experimental observations raise several important questions: How does one find an overall orbital polarization in the absence of crystal field splitting and why is there no crystal field splitting of the orbitals even under high tensile strain?
To shed insight into these questions, we performed density functional theory calculations of (LAO)$_1$/(LNO)$_1$ superlattices with the full potential linearized augmented plane waves method as implemented in the WIEN2k code~\cite{wien}. Electronic correlations beyond the generalized gradient appoximation~\cite{pbe96} of the exchange correlation potential are included within the LDA/GGA+$U$ approach~\cite{anisimov93} using $U=4$~eV and $J=1$~eV. For these calculations, the superlattices were strained laterally to the in-plane lattice parameters of STO and LAO and the out of plane parameter was set according to the experimentally obtained out-of-plane lattice parameter. While previous DFT studies~\cite{lnolaoth2,millis} have used a tetragonal setup we explored the influence of strain on the octahedral rotations by optimizing the internal parameters in a monoclinic unit cell containing 20 atoms.  Fig.~\ref{dost} shows a finite density of states at E$_{\rm F}$ for the system under compressive strain (dashed line). The two-dimensional character of these conducting sheets separated by insulating LAO layers leads to an enhanced resistivity compared to bulk LNO. In contrast, for tensile strain a gap in the total density of states (Fig.~\ref{dost}) of 0.3 eV occurs just above at the Fermi level (solid line, grey area).  This distinct behavior around the Fermi level for tensile and compressive strain is consistent with the variation in resistivity of the two samples. 

As expected, all $t_{2g}$ orbitals are occupied at the Ni-sites and there is a significant occupation of both $e_g$ orbitals (see Fig.\ \ref{pdoscdn}). This is consistent with the large covalency leading to a  $3d^8L$ ground state, where $L$ denotes a hole on the oxygen site, and thus the  Ni valence state is closer to Ni$^{2+}$ rather than the expected Ni$^{3+}$. The structural and electronic properties show a strongly  asymmetric response to strain as seen in the experiment. 
The most striking effect for tensile strain is a charge disproportionation on the Ni sublattice, Ni$^{3+}\rightarrow $Ni$^{3+\delta}$+Ni$^{3-\delta}$ with a total difference in $3d$ occupation at the Ni1 (Ni$^{3-\delta}$) and Ni2 (Ni$^{3+\delta}$) sites of  $\sim0.2e$. A similar tendency was recently reported for single-layer LNO films under tensile biaxial strain~\cite{jclno}, but in the LAO/LNO superlattice as a result of confinement the effect is strongly amplified leading to the  opening of a gap of 0.3 eV just above the Fermi level. The charge ordering goes hand in hand with bond-length disproportionation driven by a {\sl breathing} mode with two nearly regular octahedra with in-/out-of-plane cation-anion distances of 2.03/1.99\ \AA\ and  1.91/1.92\ \AA\ at the Ni1 and Ni2 sites, respectively\cite{mazin}. This is very similar to the ground-sate of charge ordered NdNiO$_3$\cite{garcia} and occurs due to a reduction in bandwidth due to bond angle distortions\cite{goodenough2}. In contrast, for compressive strain the in-plane Ni-O bond length (1.92\ \AA) is significantly shorter than the out-of-plane distance (2.00\ \AA) consistent with a volume conserving tetragonal distortion.   

The influence of strain and the interface coordination in the superattices is seen clearly in the spatial distribution of holes  shown in Fig.~\ref{pdoscdn} (right panels). The key feature  for both tensile and compressive strain is the reduced number of holes  in the $p_z$-orbital of the apical oxygen while the in-plane oxygen orbital occupancies are very similar to bulk LNO. For both strain cases there is  $\sim$6\% reduction of the apical oxygen hole density which   results from the stronger ionicity of the Al$^{3+}$-O- vs.\ Ni$^{3+}$-O-bond. To first order  this effect is larger than that of strain, which gives rise to a chemical means of controlling the orbital occupancies in addition to  strain as also noted by a recent theoretical study\cite{millis}. We note that for tensile strain the hole at the apical oxygen of Ni1 (Ni$^{3-+\delta}$) is almost completely suppressed due to the increased occupancy of the \dzt\ orbital closing the path to form the $3d^8L$ state. 

Using the insight from theory, we can see clearly that the ground state of this system is a result of the interplay of three effects: strain, confinement, and interfacial covalency. The asymmetric response to strain is reflected in the stabilization of a charge ordered state under tensile strain, which changes the unit cell volume and leads to isotropic crystal fields in the two distinct sites. On the other hand, under compressive strain the expansion of the out-of-plane Ni-O bond length results in a  crystal field splitting between the \dxtyt\ and \dzt\ orbitals and an energy lowering of the latter (see Fig.~\ref{pdoscdn}b).  This is analogous to previous results for compressive strain in thick LNO films that show quantitative agreement in the altered crystal field to the splitting of the orbitally resolved bands at the $\Gamma$ point\cite{smay}. By far though, the most profound effect is the reduction of the apical oxygen hole density. This leads to a reduction in the \dzt\  orbital occupancy under tensile strain even in the absence of crystal field splitting of the orbitals. For compressive strain, the strain induced crystal field should enhance the occupancy of the \dzt\ orbital, which is not observed. This is due to the fact that the enhanced occupancy from strain is offset by the reduction due to the reduced covalency at the Ni-O-Al bond at the interface.

In summary, the key aspect of this study is that control of orbital occupancies results from a competition of several effects: correlation, lattice distortions, chemistry and strain. These results highlight that rational manipulation of orbitals requires an understanding that extends beyond the conventional picture of how strain alters electronic properties. Chemical control offers new opportunities to control orbital configurations in ultra-thin oxide heterostructures even in the absence of strain.

\acknowledgments
Work at the Advanced Photon
Source is supported by the U.S. Department of Energy, Office of
Science under grant No. DEAC02-06CH11357. J.C. was supported by DOD-ARO under the grant
No. 0402-17291 and NSF grant No. DMR-0747808. R.P. acknowledges support by DFG (TRR80) and a grant for computational time at the Leibniz Rechenzentrum.

\end{document}